\documentclass[pss,fleqn]{w-art}
\usepackage{times}
\usepackage[]{graphicx}
\newcommand{\be}{\begin{equation}}
\newcommand{\ee}{\end{equation}}
\newcommand{\bea}{\begin{eqnarray}}
\newcommand{\bean}{\begin{eqnarray}\nonumber}
\newcommand{\eea}{\end{eqnarray}}
\newcommand{\Eq}[1]{Eq.\,(\ref{#1})}

\newcommand{\Fig}[1]{Fig.\,\ref{#1}}

\newcommand{\bfi}{\begin{figure}}
\newcommand{\efi}{\end{figure}}
\newcommand{\bmi}{\begin{minipage}}
\newcommand{\emi}{\end{minipage}}

\newcommand{\q}{{\mathbf q}}

\newcommand{\wq}{w_{\mathbf q}}

\newcommand{\ea}{\epsilon_{\alpha}}

\newcommand{\Bad}{B^{\dagger}_{\alpha}}

\newcommand{\inte}{\int\!}

\renewcommand{\dag}{^{\dagger}}

\begin{document}
\renewcommand{\copyrightyear}{2006}
\DOIsuffix{theDOIsuffix}
\Volume{XX}
\Issue{1}
\Month{01}
\Year{2006}
\pagespan{1}{4}
\Receiveddate{19 February 2006}
\Reviseddate{29 March 2006}
\Accepteddate{?}
\Dateposted{?}

\title[Memory and correlation...]{Memory and correlation effects in the exciton-phonon kinetics}
\author[G. Mannarini]{Gianandrea Mannarini\footnote{Corresponding
     author: e-mail: {\sf Gianandrea.Mannarini@unile.it}, Phone: +39\,0832\,29\,8114,
     Fax: +39\,0832\,29\,8237}\inst{1,2}} \address[\inst{1}]{National Nanotechnology Laboratory of CNR, via
Arnesano 16, 73100 Lecce, Italy}
\author[R. Zimmermann]{Roland Zimmermann\inst{2}}
\address[\inst{2}]{Institut f\"ur Physik der Humboldt-Universit\"at
   zu Berlin, Newtonstr. 15, 12489 Berlin, Germany}
\begin{abstract}
Memory and correlation effects in the interband absorption from
quantum wells due to exciton-phonon dynamics  are investigated.
They are traced back to the frequency dependence and matrix
character of the self energy arising in a 2nd Born theory. It is
found that interpeak absorption increases with respect to the
case in which this memory and correlation effects are neglected.
\end{abstract}
\maketitle

The interaction between excitonic and phonon modes in
semiconductor nanostructures leads to both relaxation of the
excitonic population as well as to pure dephasing of the optical
polarization. In small quantum dots, where the interstate
relaxation is tiny, clear signatures of pure dephasing effects
have been measured, see e.g. the extensive work done in
\cite{borri01} and \cite{borri05}. In this case, an accurate theoretical modeling is possible in the framework of an extended version of the Independent Boson Model \cite{mulj04}.  In quantum wells (QW), disorder-localized exciton
states have  significant spatial overlap and can be spectrally
rather close, thus a combination of population relaxation and
pure dephasing is expected. However, as a consequence of the
so-called Markov approximation, the absorption spectrum of a QW
results to be a superposition of Lorentzian peaks, thus
accounting for population relaxation only. We improve on this
situation by going beyond the Markov approximation in the
solution of a 2nd Born density-matrix theory  for excitons and
acoustic phonons in disordered QW.

We start here with the same model which was used in
\cite{mann06}: center-of-mass (COM) excitons are created in
disorder eigenstates $|\alpha\rangle$ of the QW by $\Bad$ and are
coupled to longitudinal acoustic phonons ($a_\q\dag$ creates a
bulk like phonon state $|\q\rangle$) via deformation potential
interaction:
\be     \label{ham} {\cal H} =  \sum_\alpha \epsilon_\alpha \, B^{\dagger}_\alpha  B_\alpha
\,        + \, \sum_{\q} \hbar\wq \, a_\q\dag a_\q
+ \sum_{\alpha \beta \q} t^\q_{\alpha\beta} (a_\q\dag + a_{-\q} )  B^{\dagger}_\alpha  B_\beta   \,.
\ee
The eigenenergies $\ea$ are obtained from a microscopic
simulation of the COM exciton in a QW with interfacial disorder
\cite{mann03} and the matrix elements $t^\q_{\alpha\beta}$ ensure
spatial overlap of the exciton states $|\alpha\rangle$ and
$|\beta\rangle$ and the phonon wavefunction $|\q\rangle$. Final
output of our theory is the linear absorption spectrum
$\alpha(\omega)$, which is connected to the Fourier transform of
the exciton polarization $P_\alpha(t)=\langle
B^\dagger_\alpha(t)\rangle$ via
\be
\alpha(\omega) = \mathrm{Im} \sum_\alpha m_\alpha P_\alpha(\omega) 
\ee
where $m_\alpha$ are matrix elements of the dipole coupling to
the light field. Both $m_\alpha$ and $t^\q_{\alpha\beta}$ are
given explicitly in \cite{mann06}.  When deriving the equation of
motion for  $P_\alpha(t)$, a hierarchy of equations is obtained
in which the polarization couples to expectation values of mixed
exciton-phonon operators. Combining  two phonon operators  into
the phonon occupation function $n(\omega_\q)=\langle a^\dagger_\q
a_\q\rangle$ allows one to end up with a  differential equation
for $P_\alpha(t)$ containing a time integration over the past and
cross correlations to all other exciton states $\beta\neq
\alpha$:
\be\label{t_domain} (-i\hbar\partial_t -
\epsilon_\alpha)P_\alpha(t)= m_\alpha  \delta(t)\, +\,
\hbar^{-1}\sum_\beta\inte dt'\, \Sigma_{\alpha\beta}(t -t')
P_\beta(t' ) \,. \ee
Here we have considered an impulsive optical excitation of
unit strength at $t=0$. The kernel $\Sigma_{\alpha\beta}(t)$
contains 2nd powers of the exciton-phonon coupling
$t^\q_{\alpha\beta}$: thus, this truncation scheme is called 2nd
Born approximation. When memory and correlation are neglected,
the kernel reduces to
\bean\label{markov}
\Sigma_{\alpha\beta}(t) &\rightarrow&  \delta_{\alpha\beta}\, \delta(t) \left[\Delta_\alpha + i \frac{\hbar\gamma^M_\alpha}{2}\right]\\
 \gamma^M_\alpha &=& \frac{2\pi}{\hbar} \sum_{\eta\q} |t_{\alpha\eta}^\q|^2 \left[ n(\omega_\q) \delta(\epsilon_\eta-\ea-\hbar\omega_\q) + \left(1+n(\omega_\q)\right) \delta(\epsilon_\eta-\ea+\hbar\omega_\q) \right]
\eea
where $\Delta_\alpha$ is a polaron shift (in the following
neglected because tiny) and $\gamma^M_\alpha$ is the Fermi's
golden rule scattering rate. The label $M$ reminds that this
further simplification is also called Markov approximation.
Clearly, this leads to an absorption spectrum which is a
superposition of Lorentzian peaks. However, finer structures are
found in the spectrum $\alpha(\omega)$ if the Markov
approximation \Eq{markov} is not carried out. For computing them,
we find convenient to Fourier transform \Eq{t_domain}, leading to
the frequency-domain version of the kernel 
\be \label{F_self} \Sigma_{\alpha\beta}(\omega) = \sum_\eta\int\!dE\,
 \frac{n(E)\,J_{\alpha\beta}^\eta(E)}{\hbar\omega -i0^+ - \epsilon_\eta + E}  \, .
\ee
The kernel or self energy $\Sigma_{\alpha\beta}(\omega)$ depends
on a coupling function $J_{\alpha\beta}^\eta(E)$ which decides
which exciton states are coupled to the phonon bath,
\be
\label{J_def} J_{\alpha\beta}^\eta(E) = \mathrm{sgn}(E)
\sum_\q  t_{\alpha\eta}^\q t_{\eta\beta}^{-\q}   \,  \delta(|E|
- \hbar\omega_\q) \,.
\ee
In particular, $J_{\alpha\beta}^\eta(E)$ is proportional to the
overlap between $|\alpha\rangle$ and $|\eta\rangle$ and between
$|\eta\rangle$ and $|\beta\rangle$. The self energy approach
leads not only to Lorentzian peaks (or Zero Phonon Lines, ZPLs),
but also to satellite absorption bands (broad bands)  in $\alpha(\omega)$. However, since
$O(N^3)$ components of $J$ have to be computed, this level of
approximation is computationally expensive when the number $N$ of
exciton states gets large. Further, it would be interesting to
have a level of approximation in which the cross-correlations
among exciton states predicted from \Eq{t_domain} are neglected,
but the memory effects are still accounted for. Thus, we consider
also a diagonal (D) non-Markov  approximation, which has got a
numerical cost $O(N^2)$  but still reproduces both Lorentzian
ZPLs and broad bands  in the
absorption with acceptable accuracy. In this case, the self
energy is diagonal, i.e. $\Sigma_{\alpha\beta}^D(\omega)=
\delta_{\alpha\beta} \Sigma_{\alpha\alpha}(\omega)$. The
differences between the spectrum $\alpha_F(\omega)$ computed using the
full self energy \Eq{F_self} and the spectrum $\alpha_D(\omega)$
obtained within the diagonal approximation are due to off diagonal
elements of the Green's function
\be\label{green} \left[G(\omega)\right]^{-1}_{\alpha\beta}  =
(\hbar\omega-i0^+-\ea)\delta_{\alpha\beta} -
\Sigma_{\alpha\beta}(\omega)  \,, \ee
which vanish within the diagonal non-Markov approximation.
\Eq{green} is derived from Fourier transformation of
\Eq{t_domain} and by noticing that $P_\alpha(\omega)= \sum_\beta
G_{\alpha\beta}(\omega) m_\beta$.

\bfi[t]
\includegraphics[width=.9\textwidth]{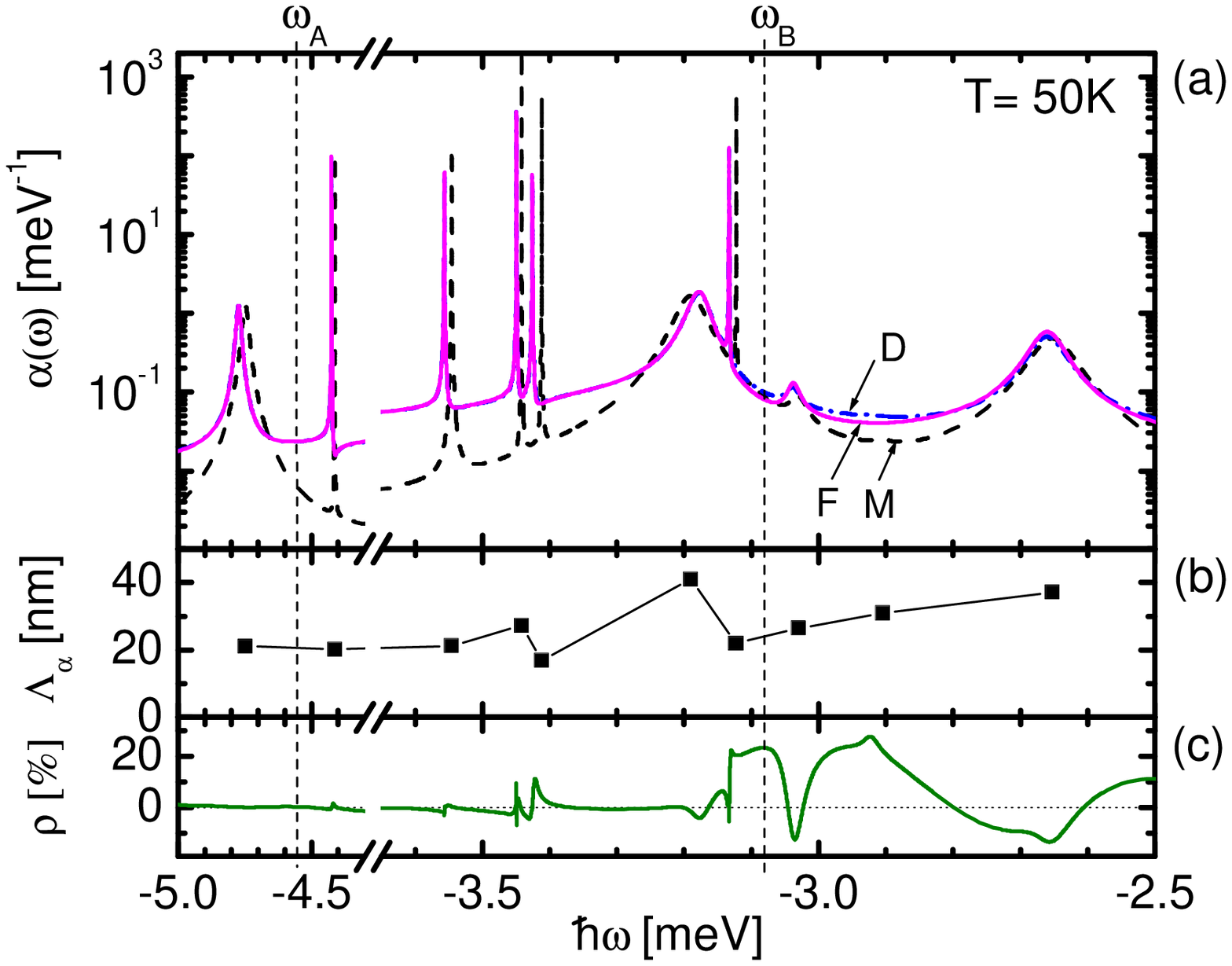}
\caption{(Color online) (a) Interband absorption spectrum
$\alpha(\omega)$ of 10 exciton states of a 5\,nm GaAs/$\mathrm{Al}_{0.3}\mathrm{Ga}_{0.7}$As QW at a lattice
temperature $T=$50\,K. The full line refers to the full (F) self
energy computation, dash-dotted is the diagonal non-Markov (D) approximation,
while dashes give the Markov (M) spectrum (superposition of
Lorentzians). The 9th peak is not resolved because its oscillator
strength $m_9$ is vanishing small. (b) Localization lengths
$\Lambda_\alpha$ of the 10 states calculated as inverse
participation ratios.  (c) Relative
accuracy $\rho=(\alpha_D-\alpha_F)/\alpha_F$. The vertical cuts
refer to the frequencies $\omega_A$ and $\omega_B$ used in
\Fig{im:fig2}.} \label{im:fig1}
\efi
For evaluating the presented theory, we consider a system of 10
exciton COM states obtained from diagonalization of a
two-dimensional Schr\"odinger equation with correlated disorder
potential, along the lines described in \cite{mann03}.  The 10 eigenfunctions
$|\alpha\rangle$  are used for evaluating the coupling functions
$J_{\alpha\beta}^\eta(E)$ and, using also the eigenenergies
$\ea$, the self energy and the absorption spectra  can be
calculated. In \Fig{im:fig1}a we compare the Markovian absorption
(superposition of Lorentzian peaks) to the spectra obtained using
non-Markovian self energies (both D and F version). The most
striking feature is the enhancement of the inter-peak absorption
with respect to the Markovian case. This effect is more important
for states which are spatially less extended, as indicated by the
localization lengths $\Lambda_\alpha$ displayed in
\Fig{im:fig1}b. The formation of the broad bands is {\em mostly}
due to the frequency dependence in the self energy denominator
(cp. \Eq{F_self}). This frequency dependence is a many-body
effect indicating that memory in the exciton-phonon kinetics is
not negligible. Since this phenomenon does not need to be related
to relaxation among different states (set all indexes equal to
$\alpha$ in \Eq{F_self}), it is also said to be an effect of
``pure dephasing''. Minor effects present in the non-Markovian
spectra (both D and F) are tiny renormalization of the peak
positions (polaron shifts) and small dispersive features in the
ZPLs: Both features are due to the real part of the self energy.
Any residual difference between the D and F spectra  is due to
the cross-correlations predicted from \Eq{t_domain}. In
\Fig{im:fig1}c we show that they are responsible for deviations
up to about 20\% in the numerical values of the broad band
absorption, while the ZPLs are mostly correctly accounted for. The
physical origin of the correlations is that the self energy
approach accounts for scattering events $|\alpha\rangle
\rightarrow |\eta\rangle \rightarrow |\beta\rangle$, where a
third state $|\eta\rangle$ can ``bridge'' between initial and
final state. Thus, it is not necessary that $|\alpha\rangle$ and $|\beta\rangle$ do overlap for scattering into each other. Due to the $t$ matrix elements appearing in \Eq{J_def}, just $|\alpha\rangle$ and $|\eta\rangle$, and $|\eta\rangle$ and $|\beta\rangle$ independently, have to overlap. 
 Conversely, within both diagonal non-Markov and
Markov approximation only one step transitions $|\alpha\rangle
\rightarrow |\beta\rangle$ are allowed in which $|\alpha\rangle$ and $|\beta\rangle$ directly overlap.
A powerful tool for checking which states are correlated is
the evaluation of the Green's function matrix
 $G_{\alpha\beta}(\omega)$. In \Fig{im:fig2} we display it
for a frequency $\omega_A$ where D and F spectra of
\Fig{im:fig1}a agree, and another frequency $\omega_B$ where
large differences occur, and correlations must be important.
Indeed a strong coupling in the subspace of states
$|6\rangle,|8\rangle,|9\rangle,|10\rangle$ is observed. The
corresponding wavefunctions exhibit large spatial overlap and $|6\rangle$ has the character of ``bridge-state'' (not shown).  
\bfi[t]
\includegraphics[width=.9\textwidth]{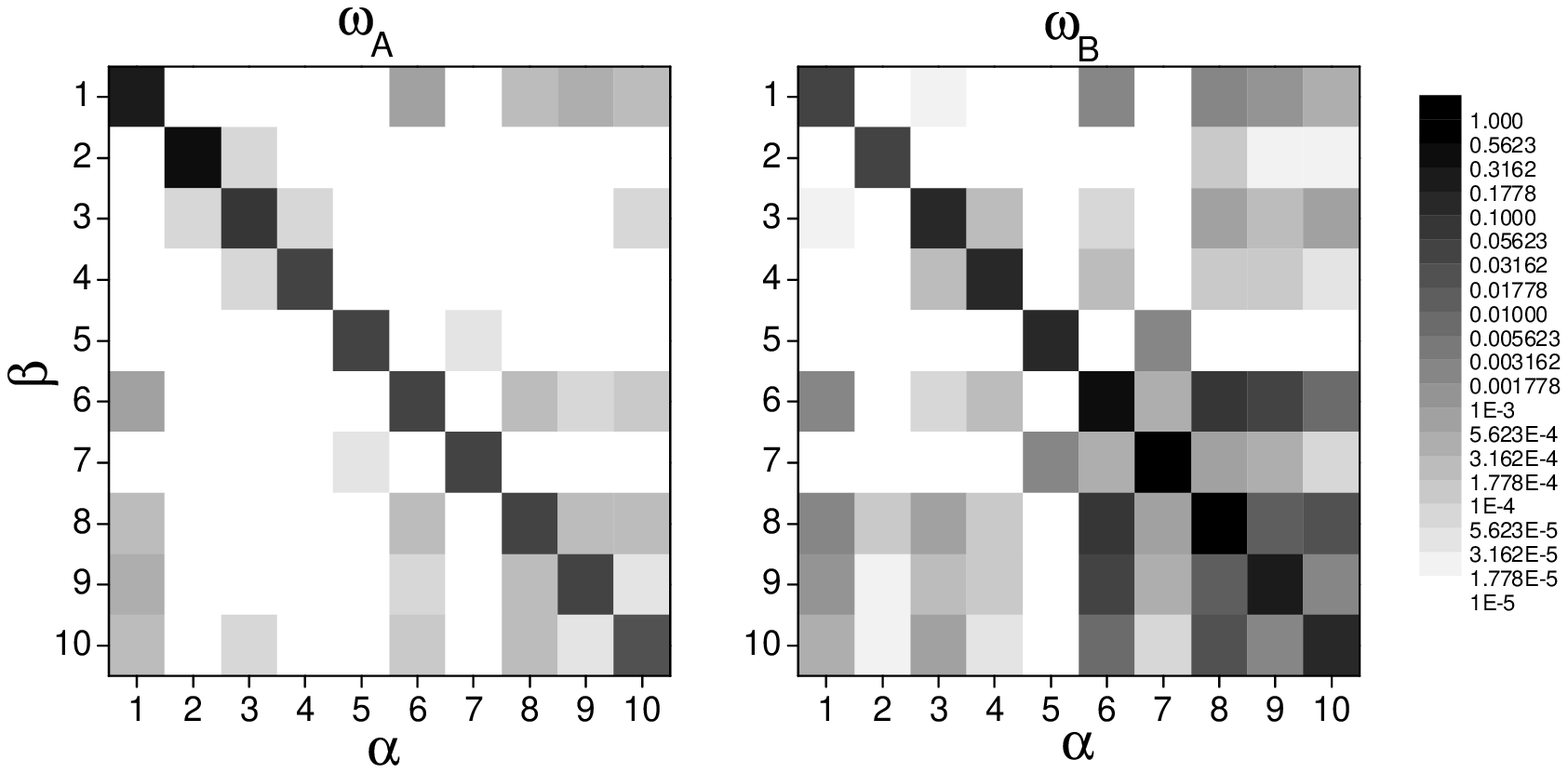}
\caption{ Green's functions
$|G_{\alpha\beta}(\omega)|$ at frequency $\omega= \omega_A$ (left
matrix) and $\omega= \omega_B$ (right matrix) defined by the cut
positions in \Fig{im:fig1}.  Logarithmic gray scale over 5 orders
of magnitude; values normalized to the maximum of
$|G_{\alpha\beta}(\omega_B)|$. For a strictly diagonal self
energy, all off diagonal elements should be zero (white).}
\label{im:fig2} 
\efi

In conclusion, we have shown that memory and correlation effects,
predicted from a non-Markovian theory of the exciton-phonon
kinetics, affect the QW absorption spectrum. Enhanced
inter-peak absorption is found using realistic wavefunctions for
a disordered QW, which is due to memory. Furthermore, two-step
scattering events lead to correlations in the optical response.
It should be possible to demonstrate these predictions by means of a QW absorption measurement,
 which became available recently \cite{guest02}.

\begin{acknowledgement}
 Support from DFG in the frame of Sfb 296 is gratefully
acknowledged. G.~M. also acknowledges support from CNR.
\end{acknowledgement}


\begin{thebibliography}{6}
\bibitem{borri01} P. Borri, W. Langbein, S. Schneider,  U. Woggon, R. L. Sellin, D. Ouyang, and D. Bimberg, Phys. Rev. Lett. {\bf 87}, 157401 (2001).
\bibitem{borri05} P. Borri,  W. Langbein,  U. Woggon,  V. Stavarache,  D. Reuter, and  A. D. Wieck, Phys. Rev. B  {\bf 71}, 115328 (2005).
\bibitem{mulj04} E. A. Muljarov and R. Zimmermann, Phys. Rev. Lett. {\bf 93}, 237401 (2004).
\bibitem{mann06} G. Mannarini and R. Zimmermann, Phys. Rev. B {\bf 73}, 115325 (2006).
\bibitem{mann03} G. Mannarini,  R. Zimmermann, G. Kocherscheidt, and W. Langbein, phys. stat. sol.(b) {\bf 238}, 494 (2003).
\bibitem{guest02} J.~R. Guest,  T. H. Stievater, Xiaoqin Li, Jun Cheng,  D. G. Steel, D. Gammon, D. S. Katzer, D. Park, C. Ell, A. Thr\"anhardt, G. Khitrova, and H. M. Gibbs , Phys. Rev. B  {\bf 65}, 241310(R) (2002).
\end{thebibliography}
\end{document}